\documentstyle[preprint,aps]{revtex}
\begin{document}
\draft
\preprint{RU9488}
\title{Hadron-nucleon Total Cross Section Fluctuations from
Hadron-nucleus\\
Total Cross Sections}
\author{David R. Harrington}
\address{Department of Physics and Astronomy, Rutgers University\\
P.O. Box 849, Piscataway, NJ 08855-0849 USA}
\date{\today}
\maketitle
\begin{abstract}
The extent to which information about fluctuations in
hadron-nucleon total cross sections in the frozen approximation
can be extracted from  very high energy hadron-nucleus total
cross section measurements for a range of heavy nuclei is
discussed.  The corrections to the predictions of Glauber theory
due to these fluctuations are calculated for several models for
the distribution functions, and differences of the order of 50 mb
are found for heavy nuclei. The generating function for the
moments of the hadron-nucleon cross section distributions can be
approximately determined from the derivatives of the hadron-nucleus
total cross sections with respect to the nuclear
geometric cross section.  The argument of the generating
function, however, it limited to the maximum value of a
dimensionless thickness function obtained at zero impact
parameter for the heaviest nuclear targets: about 1.8 for pions
and 3.0 for nucleons.
\end{abstract}
\narrowtext
\section{INTRODUCTION}

     There has recently been a revival of interest in the use of
the cross section distribution function to describe certain
features of hadronic collisions~\cite{1,2,3}.  The general approach,
which goes back to the work of Good and Walker~\cite{4} in 1960, begins
with the composite nature of hadrons and assumes that at high
energies the internal degrees of freedom are ``frozen'' during the
collision~\cite{5}.  Many features of high energy reactions can then be
discussed, at least approximately, in terms of a single
distribution function which gives the probability that the
initial hadron is in one of the configurations which interacts
with a given total cross section.  Discussions of this
distribution function for incident pions and nucleons interacting
with nucleons, including estimates of the 2nd and 3rd moments,
are given in Refs.~[1] and [2]. The discussion below tries to answer
the question of whether more information about these
distributions, and thus more information about the composite
nature of the hadrons, can be obtained by a careful study of
hadron-nucleus total cross sections.

     The answer to this question is closely related to the
importance of inelastic intermediate states in hadron-nucleus
scattering, which has been discussed extensively~\cite{3}. In
particular the systematic experimental study of Murthy {\it et al}.~\cite{6}
confirmed earlier theoretical estimates of these effects,
including their energy dependence.  Here a similar discussion is
presented in the language of the cross section distribution
function.  This approach is valid only in the high energy limit,
where the frozen approximation has a chance of being valid, but
has the advantage of including all orders of inelastic
scattering, which might well be important for heavy nuclei.

     Section II gives a brief review of how the distribution
function is defined, and what is known about it.  Then Sec.~III
summarizes the approximations and assumptions which are necessary
to obtain a simple formula for the total hadron-nucleus cross
section in terms of the generating function for the reduced
distribution function, the argument being a reduced nuclear
thickness function.  In Sec.~IV the limiting case of uniform
nuclear density is discussed. It is shown that in this case the
generating function of the distribution function is just the
derivative of the total cross section with respect to twice the
cross sectional area of the nucleus.  This result shows clearly
that the amount of information we can hope to obtain from nuclear
cross sections is limited because the argument of the generating
function is proportional to the nuclear thickness function, and
thus limited by the limited sizes of stable nuclei.  This
discussion is extended to the more realistic case of Woods-Saxon
nuclear densities in Sec.~V.  For these densities the
derivatives mentioned seem to approach the generating functions
from above for large nuclei.  In addition the numerical results
of this section  give some idea of the sensitivity of these
quantities to reasonable changes in the distribution function,
and thus an estimate of the experimental accuracy required to
distinguish among different possibilities.  These results are
discussed in Sec.~VI, along with suggestions for further
investigations in this area.

\section{Cross Section Distribution Function}

     At high energy it is assumed that the internal
configuration, here labeled by a Greek letter $\alpha$, $\beta$, etc., of a
hadron interacting with a target is ``frozen'' during the
interaction so that the transition amplitudes are diagonal in
these states:

\begin{equation}
\langle \beta ~|~F_{op}~|~ \alpha \rangle ~=~ \delta_{\beta
\alpha}F_{\alpha},
\end{equation}
where $F_{op}$ is the transition operator in the space of the internal
degrees of freedom of the incident hadron. The amplitudes for
transitions between ordinary hadronic mass eigenstates $i$ and $j$ on
a given target are then

\begin{equation}
\langle j ~|~ F_{op} ~|~ i \rangle ~=~ \sum_{\alpha} ~\langle
j~|~\alpha \rangle F_{\alpha} ~\langle \alpha ~|~ i \rangle,
\end{equation}
where $i$ and $j$ must represent states with the same flavor quantum
numbers, so $j$ must be either $i$ or a state which can be
diffractively excited from it.  Using the same normalization as
Bl\"attel {\it et al}.~\cite{1,2} the total cross section for hadron $i$
incident on the target is

\begin{equation}
\sigma_i ~=~ 4 \pi Im \langle i ~|~ F_{op} ~|~ i \rangle ~=~
\sum_{\alpha} ~ \sigma_{\alpha} ~|~ \langle \alpha ~|~ i \rangle ~|~^2,
\end{equation}
where $\sigma_{\alpha}$ is the total cross section for the hadron in
configuration $\alpha$ to interact with the target.

     If the cross section distribution function $P_i$ is defined as

\begin{equation}
P_i (\sigma) ~=~ \sum_{\alpha} ~|~ \langle \alpha ~|~ i \rangle ~|~^2
{}~\delta ~ (\sigma - \sigma_{\alpha}),
\end{equation}
this becomes

\begin{equation}
\sigma_i ~=~ \langle i ~|~ \sigma_{op} ~|~ i \rangle ~=~ ~\int~ P_i
(\sigma) ~ \sigma ~ d \sigma,
\end{equation}
i.e. $\sigma_i$ is just the average of $\sigma$ over the distribution
$P_i$.  The
same distribution can also be used to calculate the forward total
differential cross section for diffractive scattering, if the
transition amplitudes are assumed to be pure imaginary, for then

\begin{eqnarray}
\frac{d \sigma_{\rm diff}}{\rm dt} ~\mbox{{\boldmath $|$}}~{_{_{\rm
t=0}}} ~=~&& \pi
\sum_{j \neq i} ~|~ \langle j ~|~ {\rm F_{op}} ~|~ i
\rangle ~|^2 \nonumber \\
 = ~ && (\langle \sigma^2 \rangle_{\rm i} ~-~\langle \sigma
\rangle_{\rm i}^2) ~/~ (16 \pi).
\end{eqnarray}

For scattering in the high energy regime discussed by Bl\"attel
{\it et al}.~\cite{1,2}, $\langle \sigma \rangle \approx$ 24mb for pions on
nucleons while $\langle \sigma \rangle \approx$ 40mb for
nucleons on nucleons.  Experimental results on high energy
forward diffraction dissociation on nucleons gives $\langle \sigma^2\rangle
\approx$ 1.25 $\langle \sigma\rangle^2$
for nucleons, while $\langle \sigma^2\rangle\approx$ (1.4 - 1.5)
$\langle \sigma \rangle^2$
for pions. It is also
possible to put constraints on the 3rd moments of the
distributions using diffraction dissociation on deuterons, but
only by making fairly strong assumptions about the transitions
between excited states.

     For the developments below it is convenient to introduce
reduced distribution functions

\begin{equation}
f_i (x) ~ \equiv ~ \langle \sigma \rangle ~ P_i ( \langle \sigma
\rangle ~ x ),
\end{equation}
so that

\begin{equation}
\langle x^n \rangle_i ~\equiv~ \int ~ dx ~ f_i (x) ~ x^n ~=~ \langle
\sigma^n \rangle_i ~/~ \langle \sigma \rangle_i^n .
\end{equation}

The simple Poisson-like distributions

\begin{equation}
f_n (x) ~=~ [ (n+1)^{n+1} ~/~ n! ] x^n ~ {\rm exp}~ [ - (n+1) x]
\end{equation}
will be used below for illustrative purposes because they lead to
analytic formulas in some cases.  The generating function, for
example, is

\begin{equation}
\langle {\rm exp}~ (-xt) \rangle_n ~=~ \left[ \frac{(n+1)}{(n+1+t)}
\right]^{n+1} ,
\end{equation}
leading to the moments

\begin{equation}
\langle x^m \rangle_n ~=~ \frac{(n+m)~ !}{n! ~ (n+m)^m} ~ .
\end{equation}

Bl\"attel {\it et al}.~\cite{1,2} suggest several different forms for the
pion-nucleon and nucleon-nucleon cross section distribution
functions, all of which fall off with $x$ more rapidly than the
$f_n$'s.  Based on the expected behavior of the pion and nucleon
internal wave functions they also assume that f(0) vanishes for
nucleons, but not for pions .  The  behavior at large $x$, however,
is more problematic.  The main object of this paper is to see if
a careful and systematic study of hadron-nucleus total cross
section can shed any new light on these distribution functions.

\section{Hadron-Nucleus Cross Sections}

     In Glauber theory, which itself uses the frozen
approximation for the nucleons in the nucleus, the total cross
section for scattering of a hadron $i$ from a nucleus A can be
written as

\begin{equation}
\sigma_i (A) ~=~ 2 ~ \int~ d^2b \Gamma_i (A,b),
\end{equation}
where

\begin{equation}
\Gamma_i (A,b) ~=~ 1 - \langle A ~|~ \Pi_{\alpha}
{}~(1-\gamma_{i,\alpha} (b) ) ~|~ A \rangle ,
\end{equation}
with $\gamma_{\rm i,\alpha}$  the analogue of $\Gamma$ for the
scattering of the hadron $i$
from the $\alpha$th nucleon in the nucleus.  In ``standard'' Glauber
theory, where the internal degrees of freedom of the hadron are
ignored, $\gamma_{\rm i,\alpha}$ is just a numerical function of
{\bf b} and {\bf r}$_{\alpha}$, the
position of the $\alpha$th nucleon (except for possible spin- and
isospin-dependence, which we ignore here).  In general, however,
$\gamma$ should be interpreted as an operator in the internal space of
the hadron, and

\begin{equation}
\Gamma_i (A,b) ~=~ 1 - \langle A ~|~ \langle i ~|~ \Pi_{\alpha} ~
( 1 - \gamma_{\alpha op} (b) ) ~|~ i \rangle ~|~ A \rangle .
\end{equation}

This expression can be simplified considerably if a number of
assumptions are made. First, if the nuclear wave function is
completely uncorrelated

\begin{equation}
\Gamma_i (A, b) ~=~ 1 - \langle i ~|~ [1 - \langle A ~|~
\gamma_{1op} (b) ~|~ A \rangle ]^A ~|~ i \rangle .
\end{equation}

Then, if the range of $\gamma$ is much less than the nuclear size,

\begin{equation}
\Gamma_i (A,b) ~=~ 1 - \langle i ~|~ [ 1 - \frac{\sigma_{op}}{2}
T (b) ]^A ~|~ i \rangle ,
\end{equation}
where T(b), the thickness function, is the integral of the
nuclear density along a straight line at impact parameter b, and
$\sigma_{\rm op}$ is twice the integral of $\gamma_{\rm op}$ over
impact parameter space.  In
general $\gamma_{\rm op}$ and $\sigma_{\rm op}$ are both operators in
the internal space of
the hadron taking complex values, but if the imaginary part is
ignored then $\sigma_{\rm op}$ is exactly the operator described by the
distribution function $P_i$ introduced in Sec.~II.  Finally, if A
is large and $\sigma_{\rm op}$T(b) takes only small values, then

\begin{eqnarray}
\Gamma_{\rm i}~ {\rm (A,b)} ~=~&& 1 - \langle i ~|~ {\rm exp}~ [ -
\frac{\sigma_{op}}{2} {\rm T~ (b)~ A} ] ~|~ i \rangle   \nonumber \\
 =~&& 1 - \langle {\rm i} ~|~ {\rm exp}~ [ {\rm -t~ (A,b) x_{op}} ]
{}~|~ i \rangle ,
\end{eqnarray}
where

\begin{equation}
t(A,b) ~=~ AT (A,b) \langle \sigma \rangle ~/~ 2
\end{equation}
is the dimensionless reduced thickness function and

\begin{equation}
x_{op} ~=~ \sigma_{op} ~/~ \langle \sigma \rangle
\end{equation}
is the dimensionless reduced cross section operator.  In other
words, $\Gamma_i$ is determined by

\begin{equation}
\langle i ~|~ exp~ (-t ~x_{op} ) ~|~ i \rangle ~=~ \int ~ dx f_i
(x)~ exp ~ (-t~x),
\end{equation}
the generating function for the reduced distribution function
$f_i$(x).

     The reduced thickness function $t$(A,b) is expected to be
largest for b=0 and to increase nearly monotonically with A.  For
the largest stable nuclei (A=238, say) , $t_{\rm max} \approx$ 3.0 for
incident
nucleons and 1.8 for pions.  We can therefore not hope to get
information on the generating function  for $t$ greater than these
values from nuclear total cross sections.

     If $f_i$(x) does not converge rapidly enough at large $x$ then
the short-range and exponential approximations may be invalid.
The short range approximation is essential if the nuclear cross
sections are to be expressed in terms of $f_i$(x), but the
exponential approximation is to some extent just a convenience:
a version of most of the results below could be obtained without
it.  For small t

\begin{equation}
\langle i ~|~ exp~ ( -t ~ x_{op} ) ~|~ i \rangle ~=~ 1 - t + t^2
\langle i ~|~ x^2~|~ i \rangle ~/~ 2 - ... ~ ,
\end{equation}
provided the series converges.  (It does converge for all $t$ for
all the examples of Refs. [1] and [2], but for the Poisson-like
distributions the radius of convergence in n+1.)  Since $\rm \langle
i~|~ x^2
{}~|~ i\rangle$
is about 1.4 to 1.5 for pions and 1.25  for nucleons, and some
constraints on $\rm \langle i~|~ x^3~|~ i\rangle$ can be obtained from forward
diffractive
cross sections on deuterons, the generating function is already
fairly well determined for small $t$. Nuclear cross sections will
provide new information only if they can be used to constrain the
generating function for values of $t$ larger than about 1.

     The behavior of the generating function for very large $t$ is
determined by the behavior of the distribution function near x=0:

\begin{equation}
\langle i ~|~ exp ~ ( - t ~ x_{op} ) ~|~ i \rangle ~\sim ~
\frac{f_i (0)}{t} ~+~ \frac{f^{(1)}_i (0)}{t^2} ~+~ ... ~ ,
\end{equation}
where $\rm f^{(m)}_i(x)$ is the mth derivative of $\rm f_i(x)$.  If
the generating
function could be determined for large enough $t$, then, one could
check the assumptions of Refs. [1] and [2] that $\rm f_i(0)$ vanishes for
nucleons, but not for pions. Unfortunately, as noted above, the
range of $t$ is limited by the limit on the sizes of stable  nuclei
to values less than 3.

    For many purposes it is useful to decompose Eq.~(12), together
with Eq.~(17), as

\begin{equation}
\sigma_i (A) ~=~ \sigma^{(G)} _i (A) - \sigma^{(D)} _i (A),
\end{equation}
where

\begin{equation}
\sigma^{(G)} _i (A) ~=~ 2 ~\int~ d^2b ~ G^{(G)} ~ (t(A,b))
\end{equation}
is the Glauber result, and

\begin{equation}
\sigma^{(D)} _i (A) ~=~ 2 ~\int~ d^2b ~ G^{(D)} ~ (t(A,b)).
\end{equation}
is the decrease due to the dispersion in the hadron-nucleon total
cross section. (It is easy to show that $\rm G^{(D)}$ cannot be negative
and vanishes only if there are no fluctuations in $\sigma$.) The
integrand in Eq.~(24) is simply

\begin{equation}
G^{(G)} ~ (t) ~=~ 1 - ~ exp~ (-t),
\end{equation}
while

\begin{equation}
G^{(D)} ~(t) ~=~ \langle i ~|~ ~exp~ (-t ~ x_{op}) ~|~ i \rangle
{}~ - ~ exp ~ (-t).
\end{equation}

Because $\rm \sigma^{(D)}$ is a fairly small correction to $\rm
\sigma_i$(A),
in applying
these formulas to experimental data it may be sufficient to
include non-dispersive corrections, such a nuclear correlations
and corrections to the short range approximation, only in $\rm \sigma^{(G)}$,
leaving $\sigma$(D) as the simple expression above.

\section{Uniform Density Limit}

     As noted above, new information about the generating
function can be obtained from total cross sections only for heavy
nuclei.  The nuclear densities $\rho$(r) are then nearly constant in
the nuclear interior, falling quite rapidly to zero at the
nuclear surface.  In this section the extreme uniform density
limit is assumed:

\begin{equation}
\varrho (r) ~=~ \theta (R-r) ~/~ (4 \pi R^3 ~/~3),
\end{equation}
where $\rm R \approx r_0 ~ A^{1/3}$, with $\rm r_0 \approx$ 1.1 fm, is
the nuclear radius.  In this
limit, replacing A by the equivalent and more convenient label R,

\begin{equation}
t(R,b) ~=~ t(R,0) \sqrt{ 1 - (b/R)^2} ~\theta~ (R-b),
\end{equation}
where

\begin{equation}
t(R,0) ~\equiv ~ \alpha R,
\end{equation}
with

\begin{equation}
\alpha ~\equiv~ \sigma ~/~ (4 \pi r^3_0 ~/~ 3).
\end{equation}

With this simple expression for $t$(R,b), for any expression for a
contribution to a total cross section of the form

\begin{equation}
\sigma_G ~=~ 2 ~\int~ d^2b ~G(t(R,b))
\end{equation}
it is easy to show that the derivative

\begin{equation}
d \sigma_G ~/~ (2 \pi R^2) ~=~ G(t(R,0)).
\end{equation}

In other words, the slope of the cross section contribution as a
function of twice the nuclear geometric cross section is just the
function G evaluated at the maximum value of $t$(R,b) for that
nucleus, obtained at zero impact parameter.
     Since G can be chosen to be 1 minus the generating function,
or the generating function minus exp(-$t$), it is then trivial to
extract the generating function at $t$(R,0) from the derivative of
the cross section.  This relation, however, is exact only for the
uniform density case.  In the next section the accuracy of this
relation for more realistic Woods-Saxon densities is studied.

\section{Woods-Saxon Densities}

     A more realistic nuclear density, especially for the heavy
nuclei of main concern here, has the Woods-Saxon form

\begin{equation}
\varrho~ (R,r) ~=~ \frac{\varrho_0~ (R)}{1 ~+~ exp~ (( r-R) ~/~
a_0)} ,
\end{equation}
where $\rm a_0\approx$0.523 fm  is the surface thickness parameter and

\begin{equation}
1~/~ \varrho_0 ~(R) ~\approx~ \frac{4}{3} ~ \pi R^3 ~ (1 ~+~
\pi^2 ~ \frac{a_0}{R^2} )
\end{equation}
is chosen so that the volume integral of $\rho$ is unity. (The
parameter R for a number of different nuclei have been determined
by fits to total cross sections at energies below 30 GeV/c, where
inelastic contributions are expected to be small~\cite{6,7}. More
exact nuclear densities can be determined from electromagnetic
form factors, but the Woods-Saxon form is sufficient for present
purposes.)  Although there is an exact analytic expression for

\begin{equation}
t(R,0) ~=~ A \sigma \varrho_0 ~ \left\{R ~+~ \ln ~ [ 1 ~+~ exp~
(-R~/~a_0)] \right\},
\end{equation}
the thickness function for other values of b must be obtained by
numerical integration.  For this density the derivative
relationship found in Sec.~IV is not exact, although one might
expect it to be a good approximation for large R where the
density is quite uniform over most of the nucleus.  In this
section numerical results determining its accuracy are presented.

     Fig.~1 shows $\rm \sigma^D(R)$ as a function of $\rm 2 \pi R^2$
for 4 different
pion-nucleon cross section distribution functions for pion-nucleus
scattering.  Distributions (b), (c), and (d) are the same
as in Fig.~1 of Ref.~[2], while distribution (a) is the
superposition of two Poisson-like distributions introduced in
Sec.~II.  Fig.~2 shows the analogous curves for nucleon-nucleus
scattering, together with two experimental points, along with
their error bars, from Ref.~[6].  Here curve (a) is a two-term
exponential while (b) and (c) are the n=2 and n=6 distributions
shown in Fig.~4 of Ref.~[1]. (The n=10 distribution gives results
which are almost identical to those for n=6.  In using the
distributions from Refs. [1] and [2] we have adjusted the parameters
slightly so that $\langle 1\rangle~=~\langle x\rangle~= ~$1.0 more
precisely.)  For small
$\rm 2 \pi R^2$ all
the distributions give nearly the same $\rm \sigma^D(R)$ because only low
moments contribute, but for large $\rm 2 \pi R^2$ there are differences of
order 50 mb.  Since these are of the same order as the
experimental errors in Ref.~[6], it is clear that more accurate
experiments on heavy nuclei at very high energy are needed if the
hadron-nucleus cross sections are to discriminate among different
cross section distribution functions.

     In Fig.~3 the slopes in Fig.~1 are compared with the
$\rm G^D$($t$(R,0))'s
for the various distributions.  At large R  the two quantities
are similar, but $\rm G^D$ is uniformly below the derivative, although
the difference slowly decreases with increasing R.  The
corresponding results for nucleon-nucleus scattering are shown in
Fig.~4.  It seems that one might get a reasonable estimate for
$\rm G^D$ if $\rm \sigma^D(R)$ could be measured accurately enough.
This would be
particularly true if an estimate for the surface corrections to
the derivative relationship at large R could be found.

\section{Conclusion}

     The results above suggest that accurate measurements of the
total cross sections for the interactions of hadrons with a range
of heavy nuclei at very high energies could increase our
knowledge of the hadron-nucleon total cross section distribution
functions, fixing the generating functions up to fairly large
arguments and selecting among models suggested by theoretical
prejudices and constraints on the first few moments.  Very high
energies are required for the validity of the ``frozen
configuration'' approximation, while errors in measuring cross
sections ranging to more than 3000 mb must be restricted to
perhaps less than 10 mb.  The dispersive effects are larger for
pions than for nucleons, but unfortunately the maximum value of
the argument of the generating function is smaller for pions
because of their smaller cross sections on nucleons.

     Even if the results of such difficult measurements were
available, however, more work is required before they could be
unambiguously interpreted in terms of the distribution functions.
The formulas above depend upon a number of simplifying
approximations: the frozen approximation, the neglect of nuclear
correlations, the short range approximation, and the neglect of
the imaginary part of the hadron-nucleon scattering amplitude.
Fortunately, as noted in Ref.~[6], the energy dependence of the
total cross sections gives us a handle on many of these.  In the
range of energies high enough that the Glauber approximation is
valid but low enough that inelastic intermediate states do not
yet contribute significantly, the nuclear total cross sections
(as well as the differential cross sections) can be used to check
assumptions about the nuclear wave functions.  At the other
extreme, the total cross sections at very high energies should
approach constant values (aside from slow changes due to the
energy dependence of the hadron-nucleon parameters) if the frozen
approximation is valid.

     It would be useful to have more information about the
surface corrections to the relation between the generating
function and the derivative of the total cross section as a
function of R derived in Sec.~IV in the uniform density
approximation.  Preliminary investigations suggest that for
Woods-Saxon distributions these corrections should decrease as an
inverse power of R.  If this is true it should be possible to
obtain more accurate values for the generating functions by
studying the dependence of the total cross sections on R.

\acknowledgements

     The author would like to thank Harold Fearing and his
colleagues at TRIUMF, where most of the work reported here was
done, for their hospitality.  He would like to thank Brian
Jennings in particular for helpful discussions of Ref.~[3].

\begin{figure}
\caption{Dispersive contribution to pion-nucleus scattering as a
function of $\rm 2 \pi R^2$, where R is the radius parameter in the
Woods-Saxon nuclear density.  Curves b, c, and d are calculated using
the corresponding distribution functions shown in Fig.~1 of Ref.~[2],
while curve a is calculated using a distribution function
which is a superposition of the n=0 and n=1 exponential
distributions defined in Sec.~II. This distribution was chosen to
match distribution c at $\sigma$=0 and has
$\langle \sigma^2\rangle~/~\langle \sigma\rangle^2$=1.7 instead of the
value 1.5 of distributions b,c,and d.}
\end{figure}

\begin{figure}
\caption[]{Dispersive contribution to nucleon-nucleus scattering
as a function of $\rm 2 \pi R^2$, where R is the radius parameter in the
Woods-Saxon nuclear density.  Curves b, and c are calculated
using the n=2 and n=6 distribution functions shown in Fig.~4 of
Ref.~[1], while curve a is calculated using a distribution function
which is a superposition of n=1 and n=4 exponential distributions
chosen to also give $\langle \sigma^2 \rangle~=~$ 1.25 $\langle
\sigma \rangle^2$.  All these
distributions vanish at $\sigma$=0.  The experimental points with
their errors are taken from Ref.~[6].}
\end{figure}

\begin{figure}
\caption{The solid curves are the derivatives of the
corresponding curves in Fig.~1, while the adjacent dashed curves
are the $\rm G^{(D)}$ of Eq.~(27) for the same distributions, evaluated at
$t$(R,0) as given by Eq.~(30).}
\end{figure}

\begin{figure}
\caption{The solid curves are the derivatives of the
corresponding curves in Fig.~2, while the adjacent dashed curves
are the $\rm G^{(D)}$ of Eq.~(27) for the same distributions, evaluated at
$t$(R,0) as given by Eq.~(30).}
\end{figure}

\end{document}